\documentclass[pra,twocolumn]{revtex4-1}%
\usepackage[dvips]{graphicx}%
\usepackage{bm,color}
\usepackage{amsmath,amssymb,braket}
\usepackage{amsthm}
\newcommand{\ketbra}[2]{\ket{#1}\!\bra{#2}} 
\newcommand{\ave}[1]{ \langle #1  \rangle}
\def \tr{{\textrm {Tr}}}

\theoremstyle{definition}
\newtheorem{theorem}{Theorem}

\begin{document}
\title{Security against collective attacks for a continuous-variable  quantum key distribution protocol using homodyne detection and postselection}

\author{Ryo Namiki}%
\affiliation{Department of Physics, Gakushuin University, 1-5-1 Mejiro, Toshima-ku, Tokyo 171-8588, Japan}

\date{\today}
\begin{abstract} 
We analyze the security against collective attacks for a homodyne-based continuous-variable quantum key distribution protocol using binary coherent states and postselection. We derive a lower bound of the secret key rate in an asymptotic scenario, and numerically optimize it for the case of a symmetric
Gaussian channel.
\end{abstract}

\maketitle



\section{Introduction}

Quantum key distribution (QKD) enables to share a random bit sequence, usually called secret key, between two remote users, (the sender, Alice and the receiver, Bob) through an optical signal transmission and classical communications. Provided that a few physical assumptions holds, the key can be proven information theoretically secure  whenever the statistical errors in the transmission is sufficiently small. An amount of the total errors determines the length of the secret key  whether or not an eavesdropper (Eve) could have accessed to the transmission channel. There have been considerable research activity in QKD from pure theoretical aspects to  long term fieldworks \cite{rmp-QKD09,Lo2014,RMP-QKD2020RevModPhys.92.025002}. 

Continuous-variable (CV) QKD is a family of QKD protocols, in which Bob's measurement is typically associated with a set of observables that have continuous spectrum \cite{DiamantiEntropy,diam16,Laud18}. Such a measurement is usually realized by an optical  homodyne measurement or heterodyne measurement. Experimental studies suggest  CV-QKD devices are relatively easy to be incorporated into an optical fiber network, and can be operated in a high repetition rate \cite{CVFieldTestNJPSVQKD09,CVHuang:16,Kumar15,Nakazawa17,Hirano17,Eriksson2019,Zhang_2019,LopezGrande:21}. Under an asymptotic scenario, there have been a sequence of security proofs for CV-QKD protocols with various discrete modulation \cite{Zhao09,PhysRevA.97.022310,PhysRevX.9.041064,PRXQuantum.2.020325,PhysRevX.9.021059,PhysRevA.103.012412,Denys21}.

Recently, a finite-size security of a CV-QKD protocol has been established \cite{Matsuura2021}. This protocol is based on the transmission of binary coherent states, and employs both the homodyne and heterodyne measurements. The homodyne measurement is used to generate a raw key whereas it is the statistics of the heterodyne measurement that guarantees the security through monitoring a type of channel fidelity. This fidelity test gives an upper bound of a phase error rate on a virtually defined qubit-based QKD protocol. Thereby, an asymptotic key rate can be determined by a simple formula using binary entropy function for the phase error bound \cite{ShorPreskill00,Koashi09}. Since this protocol requires both the homodyne and heterodyne measurements its experimental implementation is thought to be technically demanding. Therefore, it is natural to ask whether one can make use of such a security proof for a homodyne based CV-QKD protocol. 

Incidentally, the security statement for the homodyne-based CV-QKD protocols \cite{Hirano2003,Namiki2006} is strictly limited under the Gaussian assumption \cite{Hirano16,Namiki18} and its possible refinements beyond Gaussian attacks have been open until now.


In this paper we prove an asymptotic security of  a CV-QKD protocol with binary coherent states and homodyne detection. The structure of our protocol is essentially the same as the homodyne-based CV-QKD protocols with the postselection \cite{Hirano2003,Namiki2006}. 
Our main task is to find a lower bound of the fidelity by using the homodyne statistics. Given such a bound, the security proof can be completed by repeating the method in Ref.~\cite{Matsuura2021}. 
 We also numerically determine key rates for the case of a symmetric Gaussian channel. 

This paper is organized as follows. 
In Sec.~\ref{sec2}, we  introduce our QKD protocol and outline our security proof. 
In Sec.~\ref{appTheorem}, we prove an inequality to complete the security proof of our protocol. 
 In Sec.~\ref{sectionIV}, we show numerically determined key rates when the quantum channel is a symmetric Gaussian channel. We conclude in Sec.~\ref{closing}.

\section{protocol and security proof}
 \label{sec2}

We consider a postselection protocol whose structure is basically the same as the protocols proposed in Refs.~\cite{Hirano2003,Namiki2006}. 
 Alice sends pulsed coherent states with binary phase modulation $\ket{\pm \alpha} \ (\alpha >0)$  and Bob randomly measures $\{\hat x, \hat p\}$ via homodyne detection. Our idea is to use the homodyne statistics of the  canonical quadrature pair $(\hat x, \hat p)$ for monitoring the fidelity. A sifted key is generated from pulses on which Bob measures $x$ as follows: Let us denote by $x_\text{th} \ge 0 $ the postselection threshold. Bob tells Alice the position of the pulsed signals where his $x$-measurement outcome $x$ satisfies  $|x|\ge x_\text{th}$. For each of such pulses, Alice assigns her bit values associated with the sign of modulation, whereas Bob's bit value is determined by the sign of his outcome $x$.  

In this paper we restrict our analysis for collective attacks so that Bob is supposed to receive signals come through an unknown but fixed quantum channel. We also assume  an asymptotic scenario in which a sample mean is equal to the expectation value over the density operator of individual signals.

Following \cite{Matsuura2021} we use the system index $\tilde C \to C \to B$ on the signal transmitted from Alice to Bob. Alice prepares her coherent state on an optical mode $\tilde C$, This state is transmitted to the state on Bob's optical mode $C$ through a quantum channel ${\cal E}_{\tilde C \to C}$. Bob  virtually converts the state in the optical mode $C$ into a state in a qubit system $B$ by a completely positive (CP) map ${\cal F}_{C \to B}$. This sequence is symbolically summarized as 
\begin{align}
\tilde C \overset{\cal E}{\to} C \overset{\cal F}\to B . 
\end{align}
Then, Bob's homodyne measurement process can be effectively represented by the CP map  ${\cal F}$ and a $Z$ measurement on $B$.   

For a concrete description we may use an entanglement based picture with an initial entangled state
 in a bipartite system $A {\tilde C}$ consists of  a qubit and the optical mode $\tilde C$,
 \begin{align}
\ket{\Psi}_{A \tilde C} := \frac{1}{\sqrt 2 } (\ket{0}_A\ket{\alpha }_{\tilde C} + \ket{1}_A\ket{-\alpha }_{\tilde C} ). \label{eq00}
\end{align}
As we have assumed that Eve performs a collective attack,  the system ${\tilde C}$ is sent to Bob through a quantum channel ${\cal E}$. Thereby, the state shared between Alice and Bob can be written as 
 \begin{align}
 \rho_{AC}  = {\cal I}_A \otimes {\cal E}_{\tilde C \to C } \left( \rho_{A{\tilde C}}^{(0)} 
\right)  \label{eq01},\end{align}
   where we define
$ \rho^{(0)}:=\ketbra{\Psi}{\Psi}$ and 
   ${\cal I}_A$ represents the identity map acting on Alice's qubit system $A$. 

Let us denote by $\rho_C$  a density operator in the optical mode that Bob receives  and let be  $x \in \mathbb{R}$ an outcome of Bob's  homodyne measurement on  the $x$ quadrature.  We consider the form of the CP map that converts $\rho_C$ into a possibly subnormalized density operator of a qubit system: 
\begin{align}
{\cal F} _{C \to B } (\rho_{C } )
= \int _0^\infty K_x \rho_{ C} K_x ^\dagger  dx \label{Qn2206-1}
\end{align}
with   
\begin{align}
 K_x := \sqrt {f_\text{suc}(x)} \left( \ket{0}_{B } \bra{x}_C +  \ket{1}_{B } \bra{-x}_C   \right), \label{KBC}
\end{align} where $\{ \ket{0}_{B }, \ket{1}_{B } \}$ denotes the $Z$ basis of the qubit system $B$, $\ket{x}$ denotes the homodyne basis of the $x$ quadrature, 
and 
$f_\text{suc}(x) \in [0,1]$ represents the probability that the mode conversion is carried out associated with the measurement outcomes $\pm x $.  
 In what follows we consider the form of a step function with the postselection parameter $x_\text{th} \ge 0 $,  
 \begin{align}
 f_\text{suc}(x) =  \left\{
\begin{array}{ll}
 1 , &\quad x  \ge x_\text{th}  \\
  0 , &\quad x  \in [0, x_\text{th}).
 \end{array}
\right. 
\end{align}
Note that $\cal F$ is trace decreasing if $x_\text{th} >0$. 

Now, starting with $\rho_{AC}$ in Eq.~\eqref{eq01}, Alice and Bob generate a sifted key associated with a joint $Z_A\otimes Z_B$ measurement of the final state
\begin{align}
\rho_{AB}:= {\cal I}_{A} \otimes {\cal F }_{C \to B} ( \rho _{AC}).  
\end{align}
Thereby, the probability of obtaining a  sifted bit  is give by 
\begin{align}
p_{x_\text{th}} := & \tr [\rho_{AB} ] = \tr \left( \rho_{AC} \int_{0}^\infty \! K_x ^\dagger K_ x dx \right),  
 \label{QNsift} 
\end{align}
 and  the bit error rate of the sifted key is given by
\begin{align}
 e_\text{bit} := \frac{\tr \left[\rho _{AB} \left( \ketbra{0}{0}_A \otimes \ketbra{1}{1}_{B } + \ketbra{1}{1}_A \otimes \ketbra{0}{0}_{B }\right)\right]}{p_{x_\text{th}}}.  \label{berth}
\end{align}
Here, we note that ${p_{x_\text{th}}}$ and $e_\text{bit}$ can be determined directly through our homodyne measurement of the variable $\hat x$.

Since our protocol can be seen as a protocol based on the transmission of a qubit system, the secret key rate in the asymptotic limit can be determined by estimating the error rate in the $X$ basis often called the phase error rate \cite{ShorPreskill00,Koashi09}. 
In order to give an expression of the phase error rate on the sifted key, let us start by  noting that the operator $K_x$ in Eq.~\eqref{KBC} can be written  as 
\begin{align}
 K_x = \sqrt {2f_\text{suc}(x)} \left( \ket{+}_{B } \bra{x}_C \Pi_\text{ev} +  \ket{-}_{B  }  \bra{-x}_C  \Pi_\text{od}  \right), \label{kX}
\end{align}
where the $X$ basis of Bob's  qubit is defined as 
\begin{align}
 \ket{\pm}_{B } : = \frac{1}{\sqrt 2} (\ket{0}_{B } \pm \ket{1}_{B } ) \label{pmbd}
\end{align} 
and the projection to the even (odd) photon number subspace of the optical mode is defined as  
\begin{align}
\Pi_\text{ev} :=& \sum_{n=0}^ \infty  \left( \ketbra{2n}{2n} \right)_C   \nonumber \\   
\Pi_\text{od} :=& \sum_{n=0}^ \infty \left(  \ketbra{2n+1}{2n+1}  \right)_C   . \label{ProjEvOd}
\end{align}
 The relation in Eq.~\eqref{kX} can be derived from the Eqs.~\eqref{KBC},~\eqref{pmbd},~\eqref{ProjEvOd}, and
 the following two equations: 
\begin{align}
 \bra{x}_C( \Pi_\text{ev} + \Pi_ \text{od}) =&  \bra{x}_C \nonumber \\
 \bra{x}_C( \Pi_\text{ev} - \Pi_ \text{od})=&  \bra{-x}_C,
\end{align}
where the last relation comes from the parity of the wave functions, i.e.,  $\braket{x|n} =(-1)^n \braket{-x|n} $. 

For a notation convention we may define a pair of positive operators
\begin{align}
 M_\text{ev(od) } ^\text{suc}:= 2 \int_0^\infty   f_\text{suc}(x) \Pi_\text{ev(od)} \ketbra{x}{x}_C \Pi_\text{ev(od)} dx. \label{MevodDEF}
\end{align}
With the help of Eqs.~\eqref{Qn2206-1}~and~\eqref{kX}, this gives the probability obtaining ``$+$'' or ``$-$'' outcome at the measurement on the $X$ basis as 
\begin{align}
 \ave{ + | {\cal F}_{C \to B} (\rho_C)|  +}_B =& \tr \left( \rho _C M_\text{ev} ^\text{suc} \right )\nonumber \\ 
 \ave{ - | {\cal F}_{C \to B} (\rho_C)|  -}_B =& \tr \left( \rho _C M_\text{od} ^\text{suc} \right ). \label{QnFCMC}
\end{align}
One can readily confirm that this pair can be associated with the probability $p_{x_\text{th}}$ defined in Eq.~\eqref{QNsift} 
\begin{align}
p_{x_\text{th}} =& \tr\left[ \rho_{AB} \openone_A \otimes ( \ketbra{+}{+}+\ketbra{-}{-})_B\right] \nonumber \\
=& \tr [\rho_{AC} (M_\text{ev} ^\text{suc}+ M_\text{od}^\text{suc} )], \label{QnKKMS2206}  
\end{align} 
where  $\ket{\pm}_A = \frac{1}{\sqrt 2 } (\ket{0}_A \pm \ket {1}_A )$ represents Alice's $X$ basis.

Now, let us define the phase error rate being symmetric to the bit error rate $e_\text{bit}$ in Eq.~\eqref{berth} as 
\begin{align}
 e_\text{ph} :=& \frac{\tr \left[\rho _{AB} \left( \ketbra{+}{+}_A \otimes \ketbra{-}{-}_{B } + \ketbra{-}{-}_A \otimes \ketbra{+}{+}_{B }\right)\right]}{p_{x_\text{th}}}. \label{Qn2206PhaseError} \end{align}
From Eqs.~\eqref{QnFCMC}~and~\eqref{QnKKMS2206}, 
it can be expressed in terms of qubit-mode elements on the joint system $AC$  as 
\begin{align}
 e_\text{ph} 
 =& 
\frac {\text{Tr}(\rho_{AC}M_\text{ph}^\text{suc})}{\text{Tr}[\rho_{AC}( M_\text{ev}^\text{suc}+M_\text{od}^\text{suc}) ]} 
=  \frac {\text{Tr}(\rho_{AC}M_\text{ph}^\text{suc})}{ p_ {x _\text{th}}}
, \label{PhE1}
 \end{align}
where 
the positive operator in the numerator is defined as      
\begin{align}
 M_\text{ph}^\text{suc}:= \ketbra{+}{+}_A \otimes M_\text{od}^\text{suc}+ \ketbra{-}{-}_A \otimes M_\text{ev}^\text{suc} . \label{MM21}
\end{align}
In contrast to the bit error rate $e_\text{bit}$, the phase error rate $e_\text{ph}$ cannot be observed directly in our protocol. This is because our homodyne measurement does not implement the filtering process associated with the projections $\Pi_\text{ev}$ and $\Pi_\text{od}$.  
We thus consider an upper bound of $e_\text{ph}$, which can be determined through our protocol. 

Due to  
  Corollary 2 of Ref.~\cite{Matsuura2021},  
  the following inequality holds   for any $\kappa, \gamma  \ge 0 $ and  $\beta \in \mathbb C$: 
 \begin{align}
M_\text{ph}^\text{suc}\le B_{\kappa,\gamma,\beta} 
{\openone}_{AC}  -\kappa \Pi_\text{fid}+\gamma \Pi_-  \label{phsucBkappa}
\end{align}
 where  $B_{\kappa,\gamma,\beta}$ is defined in Appendix~\ref{appB}, and 
\begin{align}
\Pi_\text{fid}:=& \left( \ketbra{0}{0} \right)_A\otimes \left( \ketbra{\beta}{\beta}  \right)_C+  \left( \ketbra{1}{1}  \right)_A \otimes \left( \ketbra{-\beta}{-\beta}  \right)_C  \nonumber  \\
\Pi_-:=&  \left( \ketbra{-}{-}  \right)_A\otimes \openone_C.   
\label{eq03}
\end{align}
  Here, $\ket{\pm \beta}$ denotes a pair of coherent states. 
From Eq.~\eqref{phsucBkappa}, the phase error rate in Eq.~\eqref{PhE1} can be bounded from above as 
\begin{align}
e_\text{ph}
\le& \frac{ B_{\kappa,\gamma,\beta}+\gamma \text{Tr}(\rho_{AC}\Pi_-) -\kappa \text{Tr}(\rho_{AC}\Pi_\text{fid})}{p_{x_\text{th}}}. \label{Ephbound00} 
 \end{align}
Here, in the numerator, the first term $B_{\kappa,\gamma,\beta}$ is independent of the state $\rho_{AC}$. The second term is determined by Alice's state preparation  or the marginal state $\tr_C(\rho_{AC})$. In fact, this term is simply calculated to be
\begin{align}
\text{Tr}(\rho_{AC}\Pi_-)= \|  \bra{-}_A \ket{\Psi}_{A\tilde C} \| ^2 =
(1-e^{-2|\alpha|^2})/2,  \label{Qnminusterm}
\end{align}  where  $\ket{\Psi}$ is defined in Eq.~\eqref{eq00}. 
It is only the last term in the numerator of Eq.~\eqref{Ephbound00} that depends on the transmission of the signal. This term is regarded as an average  fidelity for a transmission  of the binary coherent states. In fact, with the help of Eqs.~\eqref{eq00} and \eqref{eq01},  we have the following expression:  
\begin{align}
\text{Tr}(\rho_{AC}\Pi_\text{fid})=& \frac{1}{2}  \braket{ \beta | \rho_+ | \beta }    +   \frac{1}{2}  \braket{ -\beta  |\rho_- | -\beta  },  \label{opq}
\end{align}
where we define
\begin{align}
\rho_\pm :=  {\cal E}(\ketbra{\pm \alpha }{\pm \alpha }). \label{moyadayo}
\end{align}
  
In Ref.~\cite{Matsuura2021}, such a fidelity term $\text{Tr}(\rho_{AC}\Pi_\text{fid})$ is lower bounded by using the statistics of Bob's heterodyne measurement. Here, we use another lower bound based on Bob's homodyne statistics:   
\begin{align}
\text{Tr}(\rho_{AC}\Pi_\text{fid}) \ge F_0(\rho_\pm, \beta), \label{QnfdExXX}
\end{align}  
where $F_0$ will be defined in Eq.~\eqref{QnfdExRe} below.  As we will see in next section, this function $F_0$ is determined by the first and second moments of the canonical variables: 
\begin{align}
\{\tr[\rho_\pm \hat x],  \tr[\rho_\pm \hat p] ,  \tr[\rho_\pm \hat x^2],  \tr[\rho_\pm \hat p^2] \}. \label{Qnlisted}
\end{align} Therefore, the lower bound of the fidelity can be obtained from the homodyne statistics of the quadrature pair $(\hat x, \hat p)$ observed in our protocol.

From Eqs.~\eqref{Ephbound00},~\eqref{Qnminusterm},~and~\eqref{QnfdExXX} we obtain an upper bound of the phase error rate: 
\begin{align}
 e_\text{ph}\le \frac{B_{\kappa,\gamma,\beta}+\gamma (1-e^{-2|\alpha|^2})/2 -\kappa F_0(\rho_\pm, \beta)}{p_{x_\text{th}}}
 =:\tilde e . \label{QnDefiningTildeE}
\end{align}
This gives a lower bound of an asymptotic key rate \cite{ShorPreskill00,Koashi09}
\begin{align}
 G(\alpha, x _{\text{th}})= p_{x_\text{th}}  \max \left[0,   1- h(\tilde e) -h(e_\text{bit}) \right] \label{QnfdExX} 
\end{align}
where 
 $h(e)=1 -e \log_2 e - (1-e) \log_2 (1-e)$ 
is binary entropy function, and the term  $h(e_\text{bit})$ represents the portion of bits deleted to accomplish 
an ideal error correction.

Note that $G(\alpha, x_\text{th})$ works as a key rate for any  $\kappa, \gamma  \ge 0 $ and $\beta \in \mathbb C$. 
 In practical application, we may numerically optimize $(\kappa,\gamma, \beta)$ as well as $(\alpha , x_\text{th})$ to obtain a higher key rate. For the case of a phase insensitive quantum channel with the transmission $\eta> 0$, we may consider the optimization around $\beta = \sqrt \eta \alpha$ so as not to induce much degradation in the fidelity term in Eq.~\eqref{opq}. 
 Note also that  the term $h(e_\text{bit})$ in the right-hand side of Eq.~\eqref{QnfdExX} is often replaced with $ f h(e_\text{bit})$ by considering an inefficiency parameter $f \ge 1$ so as to take into account the performance of a practical error correction process.

We will determine the function $F_0$ of Eq.~\eqref{QnfdExXX} in next section. 
After that we show numerically determined key rates for a symmetric Gaussian channel in Sec.~\ref{sectionIV}. 

\section{lower bound of the fidelity} \label{appTheorem}
In order to derive the fidelity bound in Eq.~\eqref{QnfdExXX} we prove the following theorem. 
\theorem{
Let $\rho$ be a density operator and $\ket{\beta}$ be a coherent state with the amplitude $\beta \in {\mathbb C}$, i.e.,  $\hat a \ket{\beta}= \beta \ket{\beta}$. Let us define a pair of quadrature operators as  
\begin{align}
 \hat x =&(\hat a+  \hat a^\dagger) /2 \nonumber  \\ \hat p =& (\hat a - \hat a^\dagger)/(2i). \label{FidelityEQ}\end{align} 
Then, the fidelity to a coherent state $\braket{\beta| \rho| \beta}$ is bounded from below as  
\begin{align}
 \braket{\beta| \rho  |\beta}\ge\frac{3}{2}-  \text{Tr}\left[\rho\left(  (\hat x-x_\beta)^2 +(\hat p- p_ \beta )^2 \right)\right] \label{201223b1}
\end{align}
 where $x_\beta := \braket{\beta| \hat x|\beta} = \text{Re}( \beta ) $ and $ p_\beta := (\beta| \hat p | \beta)= \text{Im}( \beta )$.  
}

This theorem was essentially proven in Ref.~\cite{Namiki07}.
Note that the standard commutation relation $[\hat a, \hat a ^\dagger] =\openone$ and the normalization in Eq.~\eqref{FidelityEQ} imply  $[\hat x, \hat p]=i \openone/2$. 

\proof{
The fidelity to the vacuum state can be bounded as 
\begin{align}
 \braket{0| \rho | 0}=& \tr [ \rho \ketbra{0}{0}] = \tr[\rho ( \openone - \sum_{n=1}^\infty \ketbra{n}{n})] \nonumber \\  \ge & \tr [\rho (\openone -\sum_{n=0}^\infty n\ketbra{n}{n})] = \tr [\rho(\openone - \hat n   )], \label{tukaretayo} 
\end{align} where $\{\ket{n}\}_{n=0,1,2, \cdots}$ denotes the number basis and $\hat n $ is the number operator. This operator can be associated with the quadrature operators as 
\begin{align}
 \hat n =  \hat  a^\dagger \hat  a =(\hat x-i\hat p)(\hat x+i\hat p) = & \ \hat x^2 +\hat  p^2 +i [\hat x,\hat p] \nonumber \\ 
 = & \ \hat  x^2 +\hat p^2 -\frac{\openone}{2}. \label{mooii}
\end{align}
Substituting Eq.~\eqref{mooii} into Eq.~\eqref{tukaretayo} we have  
\begin{align}
\braket{0| \rho | 0} \ge  \frac{3}{2}- \tr[ \rho(\hat x^2 + \hat p^2 ) ]. \label{Qn201210}
\end{align}

Let us define the displacement operator  by its action to a coherent state as
\begin{align}
 D_\beta \ket{\alpha } = \ket{\alpha + \beta }. 
\end{align}
This implies 
\begin{align}
 D_\beta  \hat a D_\beta ^\dagger  =  \hat a - \beta. 
\end{align}
Then the bound Eq.~\eqref{201223b1} is obtained from Eq.~\eqref{Qn201210} by the replacement 
\begin{align}
 \rho \to D_\beta ^\dagger \rho  D_\beta. 
\end{align}
\hfill $\blacksquare$}

Now we can readily obtain the inequality in Eq.~\eqref{QnfdExXX} by using Eq.~\eqref{201223b1} on the right-hand side of Eq.~\eqref{opq}. This determines the  quantity $F_0$ in Eq.~\eqref{QnfdExXX} as  follows: 
\begin{align}
F_0(\rho_\pm, \beta):=&  \frac{3}{2}  - \frac{1}{2}\left\{ \tr [\rho_+(\hat x - x_\beta)^2] +\tr[\rho_+(\hat p - p_\beta)^2]    \right\} \nonumber   \\
& - \frac{1}{2}\left\{ \tr [\rho_-(\hat x - x_{-\beta})^2] +\tr[\rho_-(\hat p - p_{-\beta})^2]    \right\}. 
 \label{QnfdExRe}  
\end{align}
Note that this can be written as a linear combination of the quantities listed in Eq.~\eqref{Qnlisted}. Hence, the fidelity bound can be determined 
by Bob's homodyne measurements.

\section{Numerical simulation} \label{sectionIV}
In this section,  we consider a symmetric Gaussian channel and numerically determine a lower bound of the key rate. 

We assume a symmetric Gaussian channel with transmission $\eta \in [0,1]$ and the excess noise $\xi >0$. For an input of a coherent state, the  output of the channel can be written as 
\begin{align}
 {\cal E}
  (\ketbra{\alpha}{\alpha}) = \frac{2}{\pi \xi}\int_{\gamma \in \mathbb C } d^2\gamma \ketbra{\sqrt \eta \alpha+ \gamma }{ \sqrt \eta \alpha+\gamma} e^{-\frac{2|\gamma|^2}{\xi}}.\label{nenenenene}
\end{align}

In the case of $\alpha \in \mathbb R$, this gives \begin{align}
\ave{\hat x }_{\rho_\pm} =&  \tr [ \rho_\pm \hat x ]=  \pm \sqrt \eta  \alpha  \\
\ave{\hat p }_{\rho_\pm} =& \tr [\rho_\pm \hat p ]=  0 \\
\ave{\hat x^2  }_{\rho_\pm} =& \tr[ \rho_\pm \hat x^2 ]=   \eta  \alpha ^2 + \frac{1}{4}(1+\xi)  \\
\ave{\hat p^2 }_{\rho_\pm} =& \tr [ \rho_\pm \hat x^2 ]=   \frac{1}{4}(1+\xi), 
\end{align} 
where $\rho_\pm$ is defined in Eq.~\eqref{moyadayo}. Using these expressions we can write the quantity $F_0$ in Eq.~\eqref{QnfdExRe} as 
\begin{align}
 F_0 (\rho_\pm, \beta ) =& \left(  1 - \frac{ \xi }{2} \right) - \frac{1}{2} \bigl[  (\sqrt \eta \alpha - x_\beta)^2  +  p_{\beta}^2  \nonumber \\ & \quad  \qquad + (\sqrt \eta \alpha + x_{-\beta})^2 + p_{-\beta}^2    \bigr] \nonumber \\
 =& \left(  1 - \frac{ \xi }{2} \right) - (\sqrt \eta \alpha - \beta)^2, \label{QnF0Num}
 \end{align}
where we assume $\beta >0 $ and set $(x_\beta, p_\beta ) = (\beta, 0) $ in the final expression.

From Eqs.~\eqref{eq00}~-~\eqref{berth} 
~and~\eqref{nenenenene}, 
we can write 
\begin{align}
p_{x_\text{th}} 
 =& \frac{1}{2}  \text{erfc}\left( \sqrt \frac{2}{1+\xi} (x_\text{th} - \sqrt \eta \alpha ) \right) \nonumber \\ & +  \frac{1}{2}  \text{erfc}\left( \sqrt \frac{2}{1+\xi} (x_\text{th} + \sqrt \eta \alpha ) \right), \\
 e_\text{bit}=& ({2 \ p_{x_\text{th}}})^{-1} { \text{erfc}\left( \sqrt \frac{2}{1+\xi} (x_\text{th} - \sqrt \eta \alpha ) \right)},  
\end{align}
where the complementary error function is given by
\begin{align}
{\rm erfc}\bigl(x) =\frac{2}{\sqrt \pi} \int _x^\infty e^{-t^2}dt. \label{CMERF}
\end{align}
%

Now, $p_{x_\text{th}}$ and $e_\text{bit}$ are represented as a function of $(\alpha, x_\text{th},\eta, \xi)$. Then, the phase error bound $\tilde e$ in Eq.~\eqref{QnDefiningTildeE} with  $F_0$ in the form of Eq.~\eqref{QnF0Num} is determined as a function  of $(\alpha,x_\text{th},\eta, \xi)$ and $(\kappa,\gamma,\beta)$ if we set $(\beta_R, \beta_I) =(\beta,0)$ in the definition of $B_{\kappa,\gamma,\beta}$ in Appendix~\ref{appB}. Therefore, we can calculate the key rate in Eq.~\eqref{QnfdExX} for a given set of the parameters $(\alpha,x_\text{th},\eta, \xi, \kappa,\gamma,\beta)$.

\begin{figure}[!t!]
\begin{center}
\includegraphics[width=0.9\columnwidth]{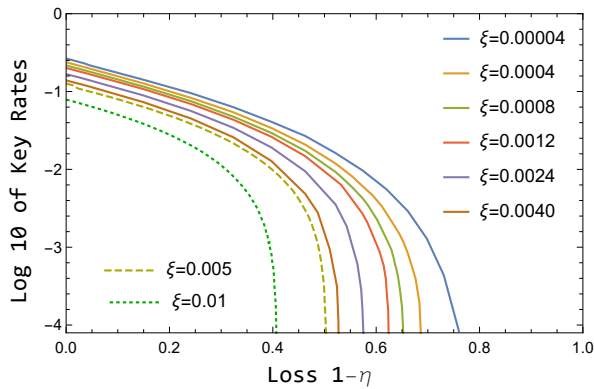}
\end{center}
\caption{Log 10 of key rates as a function of channel loss $1- \eta$.  The optimization is typically performed for $\alpha ^2 \in [0.2, 0.6]$ with a step size of $0.05$, $\gamma \in (0,2]$ with a step size of $0.02$, and $\kappa \in (0,30]$ with a step size of $0.1$. In addition, $\beta$ is somehow finely tuned around $\beta = \sqrt \eta \alpha$.  \label{keyrateFigXX}} %
\end{figure}

Figure~\ref{keyrateFigXX} shows the key rate $G$ in Eq.~\eqref{QnfdExX} as a function of the loss $1- \eta$ for a set of excess noise $\xi$. For each value of $(\eta,\xi)$, the five positive parameters $(\gamma, \kappa, \alpha,\beta, x_\text{th})$ were optimized to maximize $G$. 
Our key rate is typically lower than the key rate of four-state protocols given in \cite{PhysRevX.9.041064,PRXQuantum.2.020325} by more than one order of magnitude. 
In comparison with the two-state protocol without postselection \cite{Zhao09}, our key rates show relatively slow decay and longer transmission distances in which key rate is positive. Almost the same statement holds in conjunction with the three-state protocol without postselection  \cite{PhysRevA.97.022310}, at least in a certain parameter regime. Such a better performance in our protocol is thought to come from the use of the postselection although the difference in the security proofs potentially induces some impact. Note that the advantage of the postselection has been rather directly studied in the case of four-state protocols in Refs.~\cite{PhysRevX.9.041064,PRXQuantum.2.020325,Kani21}.


Our key rate is comparable to  the asymptotic key rate in Ref.~\cite{Matsuura2021}. As an essential technical difference to  Ref.~\cite{Matsuura2021} we use a different lower bound for the fidelity. However, it makes no much numerical significance in the present channel (See Appendix~\ref{COFB}). 
Another technical difference to Ref.~\cite{Matsuura2021}, we have considered an optimization of the parameter $\beta$, which is fixed to $\beta = \sqrt \eta \alpha $ in Ref.~\cite{Matsuura2021}. This optimization has typically induced a few percent improvement in the key rate. %

\section{Conclusion and remarks} \label{closing}

We have proven an asymptotic security of a CV-QKD protocol using binary coherent state and homodyne measurement, and develop a numerical method to show the key rate for a symmetric Gaussian channel.  The structure of our protocol is essentially the same as the CV-QKD protocols with the postselection \cite{Hirano2003,Namiki2006}, and our result would be useful to show the performance of those protocols.  
Our security proof is based on the method in  Ref.~\cite{Matsuura2021} that  established a finite size security. In contrast to Ref.~\cite{Matsuura2021} our approach uses an unbounded function for the fidelity test. This prevents us from dealing with the finite-size security at least in the present form.

For a real world application, an important aspect is to establish the security under a trusted noise scenario, in which the detection noise at Bob's station is assumed to be independent of Eve's intervention \cite{Lodewyck07,CVFieldTestNJPSVQKD09,Jougue11,Jouguet12,Us16,Namiki18,PhysRevApplied.14.064030,Laud19}. Such a direction would be addressed elsewhere.

\begin{acknowledgments}
This work was performed for Council for Science, Technology and Innovation (CSTI), Cross-ministerial Strategic Innovation Promotion Program (SIP),``Photonics and Quantum Technology for Society 5.0'' (Funding agency : QST).
\end{acknowledgments}

\appendix

\section{Definition of $B_{\kappa,\gamma,\beta}$}\label{appB}

$B_{\kappa,\gamma,\beta}$ in Eq.~\eqref{phsucBkappa} is defined through largest eigenvalues of a pair of matrices as followings: 
\begin{equation}
  \begin{split}
B_{\kappa,\gamma,\beta}  &:=  {\rm max}\left\{  \lambda_\text{max}(  M_\text{4d}^{\rm err}[\kappa,\gamma]), \lambda_{\text{max}} M_{\text{2d}}^{\rm cor}[\kappa,\gamma] ) \right\}.
  \end{split}
  \label{eq:def_of_B}
\end{equation}
Here, we use a 4-by-4 matrix,  
\begin{align}
\nonumber  &M_\text{4d}^{\rm err}[\kappa,\gamma] \\
=&    \begin{bmatrix}
        1 & \sqrt{V_{\rm od}} & 0 & 0\\
        \sqrt{V_{\rm od}} & \kappa\, C_{\rm od} + D_{\rm od} & \kappa\sqrt{C_{\rm od}\, C_{\rm ev}} & 0\\
        0 & \kappa\sqrt{C_{\rm od}\, C_{\rm ev}}, & \kappa\, C_{\rm ev} + D_{\rm ev}\!-\gamma & \sqrt{V_{\rm ev}}\\
        0 &0 & \sqrt{V_{\rm ev}} & 1 -\gamma \ 
    \end{bmatrix},
\end{align}
and a 2-by-2 matrix,  
\begin{align}
  M_{\text{2d}}^{\rm cor}[\kappa,\gamma]  &= \begin{bmatrix} \kappa\, C_{\rm ev}  & \kappa \sqrt{C_{\rm ev}\, C_{\rm od}}\  \\
        \kappa \sqrt{C_{\rm ev}\, C_{\rm od}} & \kappa\, C_{\rm od} - \gamma \end{bmatrix}.  
  \label{eq:M_cor_fin}
\end{align}
The matrix elements are specified by 
\begin{align}
    C_{\rm ev} : =& \bra{\beta}\Pi_{\rm ev}\ket{\beta} = e^{-|\beta|^2}\cosh|\beta|^2, \label{eq:definition_C}\\
    C_{\rm od} : =& \bra{\beta}\Pi_{\rm od}\ket{\beta} = e^{-|\beta|^2}\sinh|\beta|^2, \\
    \label{eq:definition_D} 
    D_{\rm ev} : = & C_{\rm ev}^{-1}\bra{\beta} M_{{\rm ev}}^{\rm suc}\ket{\beta}  \nonumber \\ =& \frac{1}{4C_{\rm ev}}\Bigl[{\rm erfc}\bigl(\sqrt{2}(x_{\rm th} - \beta_R)\bigr) + {\rm erfc}\bigl(\sqrt{2}(x_{\rm th} + \beta_R)\bigr)   \nonumber \\ &+ 2e^{-2\beta_R^2}\cos (2\beta_R\beta_I) \ {\rm erfc}\bigl(\sqrt{2}x_{\rm th}\bigr)\Bigr], \\
    D_{\rm od} : =&  C_{\rm od}^{-1}\bra{\beta} M_{{\rm od}}^{\rm suc}\ket{\beta} \nonumber \\ =& \frac{1}{4C_{\rm ev}}\Bigl[{\rm erfc}\bigl(\sqrt{2}(x_{\rm th} - \beta_R)\bigr)  + {\rm erfc}\bigl(\sqrt{2}(x_{\rm th} + \beta_R)\bigr) 
    \nonumber \\ &- 2e^{-2\beta_R^2}\cos (2\beta_R\beta_I) \ {\rm erfc}\bigl(\sqrt{2}x_{\rm th}\bigr)\Bigr],\\
    \label{eq:definition_V}
    V_{\rm ev(od)} : = & C_{\rm ev(od)}^{-1}\bra{\beta}\bigl(M_{{\rm ev(od)}}^{\rm suc}\bigr)^2\ket{\beta} - D_{\rm ev(od)}^2
     \nonumber \\ =&  D_{\rm ev(od)} - D_{\rm ev(od)}^2,
\end{align}
where the operator $M_{{\rm ev(od)}}^{\rm suc}$ and the function ${\rm erfc}(x)$ are respectively defined in Eqs.~\eqref{MevodDEF}~and~\eqref{CMERF}, and we assume the following form for the complex amplitude 
 \begin{align}
 \beta 
 = \beta_R+  i \beta_I,  \qquad \beta_R, \beta _I \in \mathbb R . 
\end{align}

\section{Comparison of fidelity bounds} \label{COFB}
\begin{figure}[!t!]
\begin{center}
\includegraphics[width=0.85\columnwidth]{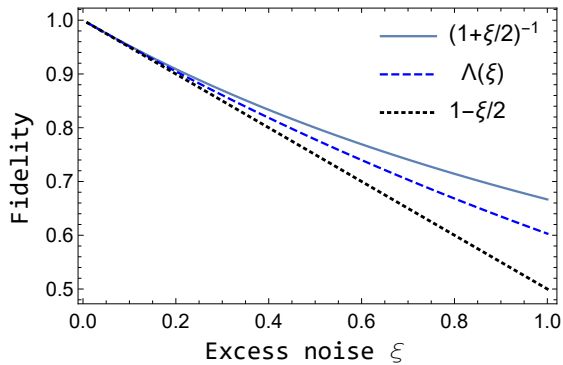}
\caption{Fidelity bounds as functions of excess noise $\xi$. Solid curve is the analytic form of  the fidelity in Eq.~\eqref{exactF}. Dashed curve is the bound defined in Eqs.~\eqref{mban1}~and~\eqref{mban2} following Ref.~\cite{Matsuura2021}. Dotted curve is our bound $1-\xi/ 2$ in Eq.~\eqref{OurBan} based on our Theorem~1. \label{FidelityFig}}
\end{center}
\end{figure}

Here, we briefly describe typical behavior of our lower bound of the fidelity and the known fidelity bound due to Ref~\cite{Matsuura2021} in the case of a symmetric Gaussian channel.
From Eq.~\eqref{nenenenene} we have the following expression to the fidelity
\begin{align}
 \braket{\sqrt \eta \alpha | {\mathcal E} (\ketbra{\alpha}{\alpha})| \sqrt \eta \alpha } = \frac{1}{1+ \xi/2}. \label{exactF}
\end{align}
Since the variances of the canonical quadrature operators for the state ${\mathcal E} (\ketbra{\alpha}{\alpha})$ are given by 
\begin{align}
 \ave{ (\Delta \hat x)^2}=  \ave{ (\Delta \hat p)^2}= \frac{1+\xi}{4}. 
\end{align}
we have the following simple form of the fidelity bound due to Eq.~\eqref{201223b1} 
\begin{align}
 \braket{\sqrt \eta \alpha | {\mathcal E} (\ketbra{\alpha}{\alpha})| \sqrt \eta \alpha }
 \ge 1-\frac{\xi}{2}.  \label{OurBan}
\end{align}
Due to Ref.~\cite{Matsuura2021}, the heterodyne statistics gives the following fidelity bound  
\begin{align}
 \braket{\sqrt \eta \alpha | {\mathcal E} (\ketbra{\alpha}{\alpha})| \sqrt \eta \alpha }
 \ge \ave{\Lambda_{m,r}}(\xi),  
\end{align}
where 
\begin{align}
\ave{\Lambda_{m,r}}(\xi) : = \frac{1}{1+ {\xi }/{2}} \left[ {1-(-1)^{m+1} \left(\frac{\xi /2 }{ 1+r (1+ {\xi }/{2})   }\right)^{m+1}} \right] \label{mban1}
\end{align}
with 
\begin{align}
 (m,r)=(1,0.4120).  \label{mban2}
\end{align}
Figure~\ref{FidelityFig} shows  the fidelity and bounds as functions of the excess noise $\xi$. There are no impact difference in the three functions when the excess noise is not significantly high, e.g., $\xi\le 0.2 $. 
In fact, our bound $1 -\xi/2$ is equivalent to the fidelity up to the second order of $\xi $, and the following relation holds:
\begin{align}
 \frac{1}{1+ \xi/2} \ge \ave{\Lambda_{m,r}}(\xi)  \ge 1-\xi/2 .
\end{align}


%

\end{document}